\documentclass{svproc}
\let\svthefootnote\thefootnote
\usepackage{url}

\usepackage{graphicx}
 \usepackage{amssymb}
 \usepackage{amsmath}

\begin{document}
\mainmatter              
\title{Exclusion Limits on Hidden-Photon Dark Matter near 2 neV from a Fixed-Frequency Superconducting Lumped-Element Resonator}
\titlerunning{Exclusion Limits on Hidden-Photon Dark Matter near 2 neV...}  
%
\author{A. Phipps\inst{1},
S. E. Kuenstner\inst{1},
S. Chaudhuri\inst{1},
C. S. Dawson\inst{1},  
B. A. Young\inst{2},\newline
C. T. FitzGerald\inst{2},
H. Froland\inst{1},
K. Wells\inst{1},
D. Li\inst{3},
H. M. Cho\inst{3},
S. Rajendran\inst{4},
P. W. Graham\inst{1}, and
K. D. Irwin\inst{1,3}
}
\authorrunning{A. Phipps et al.} 
%
\tocauthor{A. Phipps et al.}
\institute{Department of Physics, Stanford University, Stanford, California 94305, USA
\and
Department of Physics, Santa Clara University, Santa Clara, California 95053, USA
\and
Technology Innovation Directorate, SLAC National Accelerator Laboratory, Menlo Park, California 94025, USA
\and
Berkeley Center for Theoretical Physics, Department of Physics, University of California, Berkeley, California 94720, USA}

\maketitle              

\begin{abstract}
We present the design and performance of a simple fixed-frequency superconducting lumped-element resonator developed for axion and hidden photon dark matter detection. A rectangular NbTi inductor was coupled to a Nb-coated sapphire capacitor and immersed in liquid helium within a superconducting shield. The resonator was transformer-coupled to a DC SQUID for readout. We measured a quality factor of $\sim$40,000 at the resonant frequency of 492.027 kHz and set a simple exclusion limit on $\sim$2 neV hidden photons with kinetic mixing angle $\varepsilon\gtrsim1.5\times10^{-9}$ based on 5.14 hours of integrated noise. This test device informs the development of the Dark Matter Radio, a tunable superconducting lumped-element resonator which will search for axions and hidden photons over the 100 Hz to 300 MHz frequency range.    
\end{abstract}
\section{Introduction}
The Dark Matter Radio (DM Radio)\cite{dmradio,dmradiopath} is a new experiment to search for ultralight, wavelike dark-matter --- particles with individual mass $m_0 \ll 1$ eV$/c^2$. If these particles account for a substantial fraction of the cold dark matter, they must be bosonic with a high occupation number. As a consequence, this type of dark matter behaves like a classical, oscillating wave with a narrow bandwidth near frequency $f=m_0c^2/h$. Rather than search for the scattering or absorption of a single dark-matter particle, we search for a persistent, narrow-band signal caused by weak coupling of the dark matter field to Standard Model particles.\let\thefootnote\relax\footnote{To appear in Proceedings of the 3rd International Workshop on Microwave Cavities and Detectors for Axion Research}
\addtocounter{footnote}{-1}\let\thefootnote\svthefootnote    

The QCD axion\cite{peccei77,weinberg78,wilczek78} is a strongly motivated candidate dark-matter wave. Axion haloscopes\cite{sikivie83} use the inverse Primakoff effect to search for the conversion of axion dark matter into photons by immersing a tunable microwave cavity in a strong DC magnetic field. If the cavity is tuned to match the frequency of the axion, the signal rings up the cavity, which leads to a power excess above the thermal noise. The sensitivity increases with resonator volume, quality factor, and magnetic-field strength. ADMX\cite{admx2018} and HAYSTAC\cite{haystac2018} have used this technique to place constraints on the axion-photon coupling $g_{a\gamma\gamma}$ for axion masses $\sim$1 $\mu$eV$/c^2$ and higher.

Lumped-element resonators can also be used to search for the ``hidden-sector photon'', a spin-1 vector dark-matter wave, which can be produced in the observed dark-matter abundance by inflationary fluctuations\cite{graham2016} or the misalignment mechanism\cite{arias2012}. Kinetic mixing results in the conversion of hidden photons to ordinary photons. An external magnetic field is not required. Enhancement of the signal by a suitably-tuned resonator also occurs, allowing axion haloscopes to place constraints on the kinetic mixing angle $\varepsilon$ for $\sim>$1$\mu$eV hidden photons.

Cavity haloscope searches are only sensitive to axions/hidden photons whose Compton wavelength is comparable to the size of the cavity, limiting practical searches to $m_0 \gtrapprox 1\mu$eV$/c^2$. Lumped-element resonators (a wire-wound inductor connected to a monolithic capacitor) are free from this geometric restriction, motivating their use when the dark matter Compton wavelength is much larger than the size of the detector \cite{dmradio,cabrera2010,sik2014,kahn2016}. An optimized single-pole, lumped-element resonator can theoretically achieve $\sim$70\% of the fundamental quantum limit on detection of axion and hidden photon dark matter in the sub-wavelength regime\cite{fundlimits}. Tunable lumped-element resonators with a high quality factor will be required to cover QCD axion parameter space with $m_0 < 1\mu$eV$/c^2$.  

DM Radio will consist of a tunable, superconducting, lumped-element resonator designed to search for axion and hidden photon dark matter over the $\sim$peV to $\sim\mu$eV (100 Hz to 300 MHz) mass range. In this paper, we describe the performance of a simple fixed-frequency, superconducting lumped-element test resonator, and compute the first direct detection limits on hidden photons in a narrow range around 2.035 neV.

\section{Experimental Setup}
\begin{figure}
  \centering
    \includegraphics[width=\textwidth]{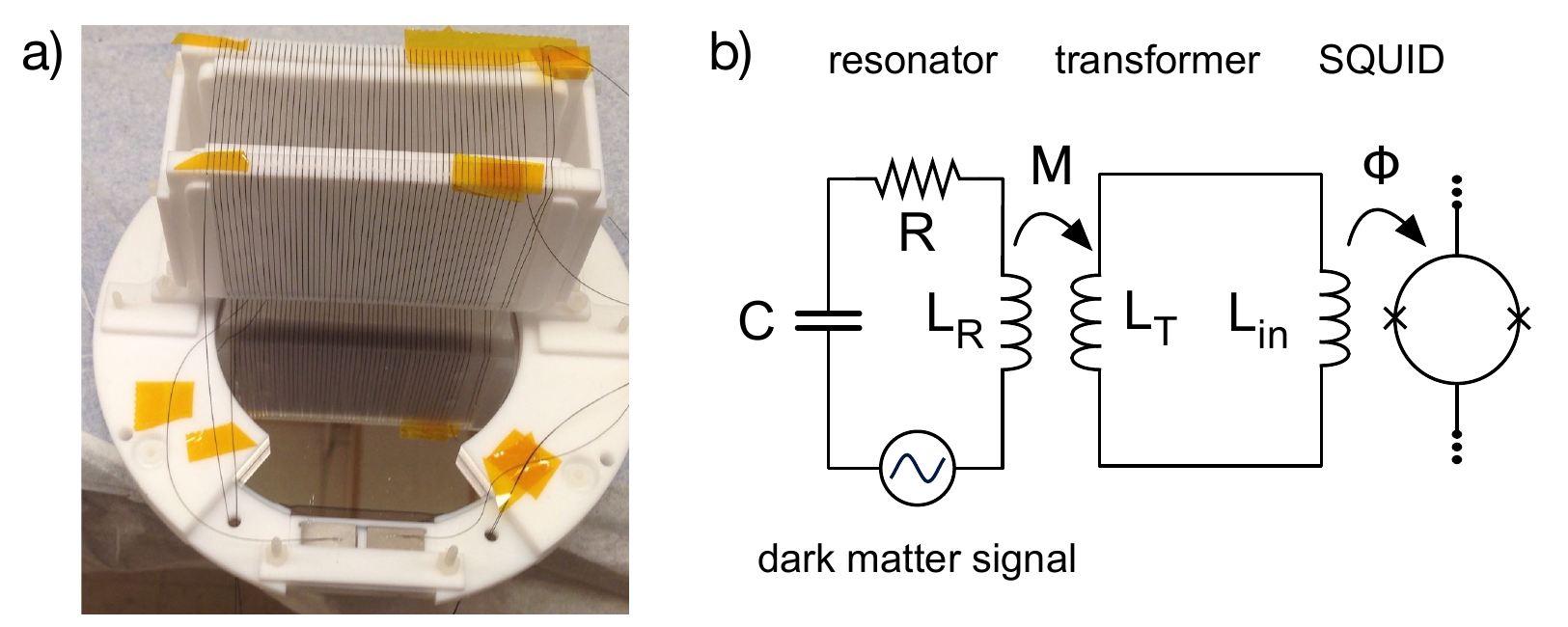}
     \caption{a) The lumped-element resonator. b) The  equivalent circuit model.}
     \label{fig:fixedres}
\end{figure}

The resonator used in this experiment is shown in Fig~\ref{fig:fixedres}a. The inductor is a 40-turn (0.05 inch pitch) Formvar-coated rectangular niobium-titanium (NbTi) coil wound on a polytetrafluoroethylene (PTFE) form, enclosing a volume of $\sim$100 mL. Opposite faces of a 500 $\mu$m thick, 100 mm diameter sapphire wafer were coated with 800 nm of Nb to form the capacitor. Electrical connections were created by Nb wirebonds between the capacitor electrodes and small Nb blocks. The ends of the inductor wire were also spot-welded to the blocks. 

A single-turn NbTi coil on the end of the PTFE form was used to inductively couple to the resonator. The ends of the transformer coil passed through a small hole into a separate Nb annex which houses a custom DC SQUID. The transformer coil is connected to two Nb screw terminal blocks using Nb screws and washers. The input coil of the SQUID was connected to the blocks using Nb wirebonds. 

The resonator and annex were enclosed in a 21.6 cm tall, 14.4 cm diameter cylindrical Nb shield with 2 mm wall thickness. The shield was mounted to a stainless steel probe and cooled to 4.2 Kelvin by insertion into a liquid helium dewar lined with Cryoperm-10. Wiring in the center tube of the probe connected the SQUID to room-temperature electronics. Biasing and pre-amplification were provided by Magnicon XXF-1 electronics. The SQUID was operated open-loop with 3.5 MHz bandwidth. Additional amplification was provided by two Stanford Research Systems SR560 channels before digitization.

The equivalent circuit model of this resonator is shown in Fig.~\ref{fig:fixedres}b. Dedicated calibration runs were performed to measure the inductance of the resonator ($L_\mathrm{R}=59\pm6$ $\mu$H), transformer ($L_\mathrm{T}=842\pm29$ nH), their mutual inductance ($M=1.15\pm0.06$ $\mu$H) and the SQUID input coil inductance ($L_\mathrm{in}=2.74\pm 0.06$ $\mu$H). The measured resonant frequency of 492.027 kHz was used to find the total capacitance ($C=1.78 \pm0.18$ nF). The resonator had a quality factor of $\sim$40,000 as determined by a Lorentzian fit to the thermal noise peak (see Fig.~\ref{fig:frpeakdistro}a), implying an equivalent series resistance of approximately 4.5 mOhm. 

A known AC current was injected into a separate inductor coil coupled to the SQUID while recording the output voltage, providing a calibration of the system gain. This measurement determines the SQUID input coil sensitivity using known values for the SQUID couplings. 

\section{Expected Signal-to-Noise Ratio}
In the sub-wavelength limit, hidden photons can be treated as isotropic, oscillating effective current density $\vec{J}_{\mathrm{HP}}$ which fills the superconducting shield\cite{dmradio}. If we assume that the effective hidden-photon current points along the longitudinal axis of the shield and ignoring their small velocity, this current takes the form
\begin{equation}
\vec{J}_{\mathrm{HP}} = -\varepsilon \epsilon_0 \left(\frac{mc^2}{\hbar}\right)^2\left(\frac{\hbar}{mc}\right)\sqrt{2\mu_0 \rho_{DM}}\exp\left(\frac{imc^2t}{\hbar}\right)\hat{z},
\end{equation}
where $\varepsilon$ is the kinetic mixing angle, $m$ is the hidden photon mass, and $\rho_{\mathrm{DM}}$ is the local dark matter density. In contrast to microwave cavities, the electric field inside the shield is highly suppressed in the sub-wavelength regime and magnetoquasistatics apply \cite{dmradio}. $\vec{J}_{\mathrm{HP}}$ sources an oscillating, quasistatic magnetic field inside the shield, which is sensed by the resonator inductor coil. The induced EMF was found by Faraday's law to be
\begin{equation}
V_{\mathrm{HP}} = \frac{i\varepsilon}{c}\left(\frac{mc^2}{\hbar}\right)^2\sqrt{2\mu_0\rho_{DM}}\exp\left({\frac{imc^2t}{\hbar}}\right)\left(\frac{Nh}{4}\left(y_2^2-y_1^2\right)\right),    
\end{equation}
where $N$ is the number of coil turns, $h$ is the height of the rectangular coil, and ($y_1$, $y_2$) denote the horizontal offset of the coil edges to the center of the shield.

The SQUID input-coil current due to a voltage $V$ in series with the resonator inductance is given by
\begin{equation}
\label{eq:iin}
i_{\mathrm{in}} = \frac{M}{L_\mathrm{T}+L_{\mathrm{in}}}\frac{V}{Z_r},
\end{equation}
where $Z_r$ is the impedance of a series RLC resonator with $L=L_{\mathrm{eff}}=L_\mathrm{R}-M^2/(L_{\mathrm{T}}+L_{\mathrm{in}})$ due to screening by the transformer. Eq.~\ref{eq:iin} was used to calculate the input-coil current spectral density $i_{\mathrm{HP}}$ due to an assumed $V_{\mathrm{HP}}$. 

Loss mechanisms in the resonator (represented by the resistance $R$) produce a Johnson-Nyquist noise voltage density $V_{\mathrm{th}}=\sqrt{4kT_{\mathrm{N}}R}$ in series with the resonator inductance. The effective noise temperature $T_\mathrm{N}$ determines the height of the input-referred thermal noise peak, while $R$ determines the bandwidth. We denote the input-coil current noise due to this thermal noise as $i_{\mathrm{th}}$. The SQUID itself adds an additional input-referred white noise $i_{\mathrm{w}}$. We fit the input-referred average power spectral density (PSD) to the uncorrelated sum of $i_\mathrm{th}$ and $i_w$ (Fig.~\ref{fig:frpeakdistro}a) which gave $R=4.53$ mOhm, $T_\mathrm{N}=9.34$ K, and $i_\mathrm{w}=1.41$ pA/$\sqrt{\mathrm{Hz}}$. 
Using these parameters, the input-referred expected signal-to-noise ratio, $\eta$, is:

\begin{equation}
\label{eq:snr}
\eta = \frac{|i_{\mathrm{HP}}|^2}{|i_{\mathrm{th}}|^2+|i_{\mathrm{w}}|^2}.
\end{equation}

\section{Data Collection and Analysis}
The test data was recorded on the morning of July 28, 2018. The amplified SQUID output voltage was sampled at 25 MHz with a 16-bit digitizer. The average PSD was formed from 6900 independent 2.68 second time records ($2^{26}$ samples, $\Delta f=0.37$ Hz), equal to a total integration time of $\tau=5.14$ hours. A Blackman-Harris (BH) window was applied to each time trace prior to PSD computation.

\begin{figure}
  \centering
    \includegraphics[width=\textwidth]{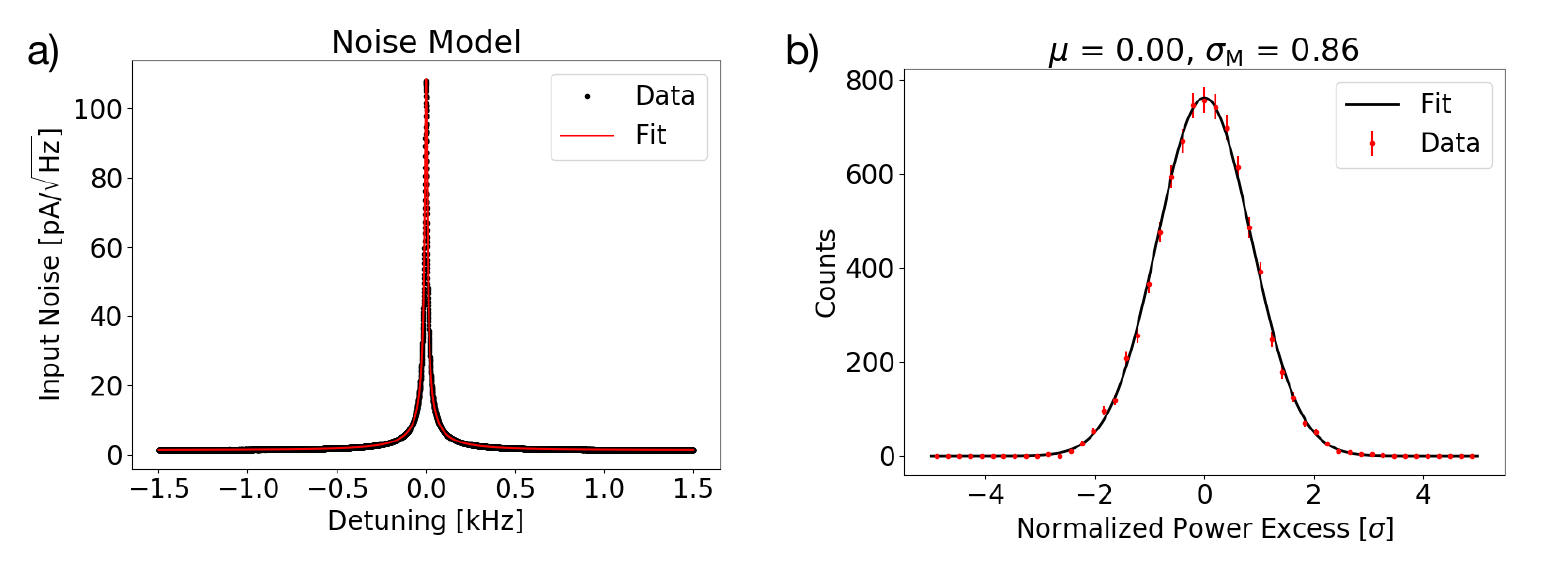}
     \caption{a) Noise model fit to the thermal peak (see text). 
     b) Histogram of the excess power distribution. Data points are shown at the bin centers, with $\sqrt{N}$ error bars.}
     \label{fig:frpeakdistro}
\end{figure}

Analysis of the data loosely followed the procedure outlined in \cite{haystacanalysis}. The spectral baseline was removed by applying a Savitzky-Golay (SG) filter (W=50, d=6) to the average PSD. The PSD and SG output were truncated to an analysis band of $492.027 \pm 1.5$ kHz. The residual excess power in each frequency bin was determined by normalizing the PSD to the SG output and subtracting 1. In the absence of a signal, each resulting bin should be an independent sample drawn from a Gaussian distribution with $\mu=0$ and $\sigma=1/\sqrt{\tau\Delta f}$. A histogram of the excess power data is shown in Fig.~\ref{fig:frpeakdistro}b. While the distribution has a mean of zero, the width is narrower than expected ($\sigma_M=0.86$). We can attribute this narrowing to the combined effects of the BH window and SG filter, which was confirmed by a Monte Carlo simulation in which the excess power distribution was formed from computer-generated noise traces with and without application of windowing/filtering. We conclude that the distribution of excess power over frequency bins was consistent with noise. In future analyses, a standard rectangular window will be used. 

\begin{figure}
  \centering
    \includegraphics[width=\textwidth]{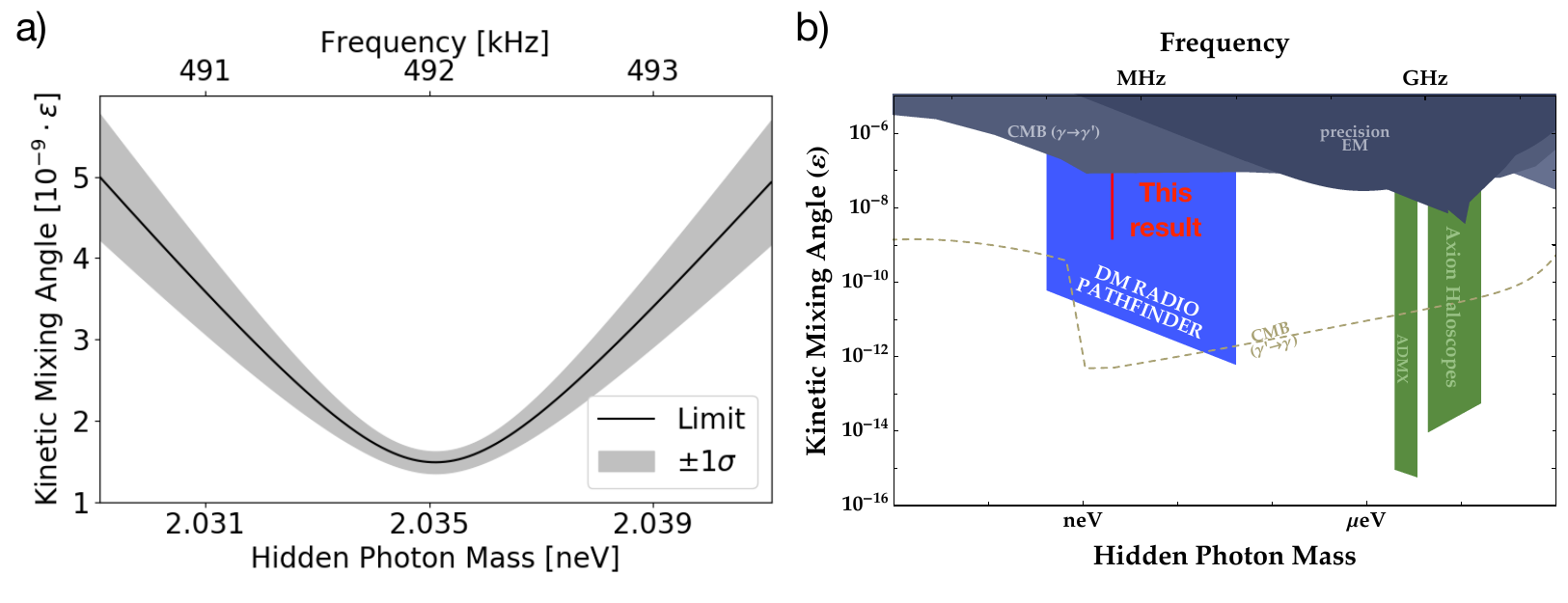}
     \caption{a) The 90\% C.L. exclusion line on hidden photon dark matter (see text)
     with $\pm1\sigma$ band due to systematic error. The exclusion is calculated assuming that the effective hidden-photon current points along the longitudinal axis of the shield; field disalignment would weaken this limit. b) Wider view of hidden photon parameter space. The result presented here is shown in red. Projected limits from a 1-year scan of the DM Radio Pathfinder are shown in blue. Limits from axion haloscopes are shown in green. The dashed line is a model-dependent constraint above which hidden photons would not account for the total dark matter density.\cite{arias2012}}
     \label{fig:hplimit}
\end{figure}

We set an upper limit on the hidden-photon kinetic-mixing angle $\varepsilon$ (normalized to a local dark matter density of $\rho_\text{DM}=0.45$ GeV/$c^2$) using the maximum observed single-bin excess power of 3.34$\sigma$. The exclusion is calculated assuming that the effective hidden-photon current points along the longitudinal axis of the shield; field disalignment would weaken this limit. Three orthogonal experiments could be used to eliminate this assumption, or multiple rescans with one experiment could be used to set a more rigorous limit with a more general assumption about directionality. We divide this value by $\sigma_M=0.86$ to correct for the effects of the BH window and SG filter. While we only performed a single-bin search for excess power, the standard halo model predicts greater than 45\% of the total signal power would be contained within a single bin. We divide our single-bin excess value by this as a correction. For each frequency bin, we use Eq.~\ref{eq:snr} to determine the kinetic mixing angle which would produce an excess power above 3.34$\sigma$/(0.86$\times$0.45)=8.63$\sigma$ at 90\% confidence. The result is shown in Fig.~\ref{fig:hplimit}(a), excluding kinetic mixing angles of $\varepsilon \gtrsim 1.5\times10^{-9}$ in a band around 2.035 neV. We note that even for frequency bins detuned by several resonator bandwidths, the exclusion limit is not significantly degraded due to out-of-band sensitivity.\cite{fundlimits}   

\section{Discussion and Conclusion}
The resonator described here was designed as a test structure for DM Radio. It lacks the physical volume, frequency scanning, and maturity of modern axion haloscopes, and the exclusion limit it demonstrates is not competitive with larger experiments (although it is in a new mass regime). This work is just a first step towards using lumped-element resonators to search for dark-matter waves in the sub-$\mu$eV regime. The resonator quality factor of $\sim$40,000, already comparable to copper microwave cavities, is believed to be limited by a combination of the low characteristic resonator impedance and loss in the sapphire capacitor. Operation in liquid helium with a non-optimized SQUID is responsible for the high noise temperature of $9.34$ K. Even with these limitations, a 5.14-hour integration was able to set an exclusion limit on the kinetic mixing angle of $\sim$2.035 neV hidden photon dark matter and is the first direct detection limit in this mass range.

 The next step in the DM Radio program is the
 DM Radio Pathfinder\cite{dmradiopath}, a tunable resonator with 7x larger volume and 10x higher characteristic impedance. 
 Projected hidden-photon limits from a 1-year scan of the Pathfinder are shown in Fig.~\ref{fig:hplimit}b. The full DM Radio will include a DC magnetic field to search for axions, with an ultimate goal of probing the QCD band in the $10 $ neV$/c^2$ -- 1 $\mu$eV$/c^2$ mass range, and a broader range of axion-like-particles and hidden photons across peV -- $\mu$eV$/c^2$ masses.

\section{Acknowledgements}

This research was sponsored by a seed grant from the Kavli Institute for Particle Astrophysics and Cosmology, by the Laboratory Directed Research and Development Program of the SLAC National Accelerator Laboratory, and by DOE HEP QuantISED award \#100495.
%
%

\end{document}